\documentclass[12pt]{iopart}
\usepackage{fullpage,graphicx,epsf}
\usepackage{iopams}
\usepackage{amssymb}
\usepackage{color,ulem}
\usepackage{amscd}
\usepackage{cite}
\usepackage{color}
\usepackage{graphicx}
\usepackage{epsfig}
\usepackage[english]{babel}
\usepackage{amsfonts}
\usepackage{verbatim}
\begin{document}
\def\be{\begin{equation}}
\def\ee{\end{equation}}
\def\bea{\begin{eqnarray}}
\def\eea{\end{eqnarray}}
\def\bml{\begin{mathletters}}
\def\eml{\end{mathletters}}
\def\l{\label}
\def\b{\bullet}
\def\no{\nonumber}
\def\fr{\frac}
\def\th{\theta}
\def\eps{\epsilon}
\def\o{\omega}
\def\O{\Omega}
\def\p{\partial}
\def\ph{\phi}
\title{Classical Heisenberg spins with long-range interactions:
Relaxation to equilibrium for finite systems}
\author{Julien Barr\'{e}$^1$ and Shamik Gupta$^2$}
\address{$^1$Laboratoire J. A. Dieudonn\'{e}, Universit\'{e} de
Nice-Sophia Antipolis, UMR CNRS
7351, Parc Valrose, F-06108 Nice Cedex 02, France \\
$^2$Laboratoire de Physique Th\'{e}orique et Mod\`{e}les
Statistiques, UMR CNRS 8626, Universit\'{e} Paris-Sud, Orsay, France}
\ead{julien.barre@unice.fr,shamikg1@gmail.com}
\begin{abstract}
Systems with long-range interactions often relax towards statistical
equilibrium over timescales that diverge with $N$, the number of particles.
A recent work [S. Gupta and D. Mukamel, J.
Stat. Mech.: Theory Exp. P03015 (2011)] analyzed a model system
comprising 
$N$ globally coupled classical Heisenberg spins and evolving under classical spin dynamics. 
It was numerically shown to relax to
equilibrium over a time that scales superlinearly with $N$. Here, we present a detailed study of the Lenard-Balescu operator that accounts at leading order for the finite-$N$ effects driving this relaxation. We demonstrate that corrections at this order are identically
zero, so that relaxation occurs over a time longer than of order $N$, in
agreement with the reported numerical results.
\end{abstract}
\date{\today}
\maketitle
\tableofcontents

\section{Introduction}

Long-range interacting systems are characterized by an interparticle
potential with a range that is of the order of the system size.
In $d$ dimensions, this corresponds to potentials decaying at large separation, $r$, as
$1/r^{\alpha}$, where $\alpha$ lies in the range $0 \le \alpha \le d$ 
\cite{review1,review2}. Examples include gravitational systems \cite{Chavanis:2006}, plasmas
\cite{Escande:2010},  two-dimensional hydrodynamics
\cite{Bouchet:2012}, charged and dipolar systems \cite{Bramwell:2010},
and many others. 

Despite obvious differences, these systems often share a common
phenomenology (see \cite{review1,review2}). In particular, the
relaxation to equilibrium of an isolated long-range interacting system
of $N$ particles proceeds in two steps: first, a collisionless
relaxation, described by a Vlasov-type equation, brings the system
close to a nonequilibrium state, called the "quasistationary state"
(QSS), whose lifetime increases with $N$; second, on timescales
diverging with $N$, discreteness effects due to finite value of $N$
drive the system towards Boltzmann-Gibbs equilibrium. This second step
is usually described by a Lenard-Balescu-type equation
\cite{Lenard:1960,Balescu:1960}. This scenario
is well established for plasmas and self-gravitating
systems\cite{BinneyTremaine}, and has been studied in detail in
various toy models for long-range interactions
\cite{Ruffo:1995,Nobre:2003,Yamaguchi:2004,Jain:2007,Gupta:2013}. The
QSS lifetime, which may be regarded to be of the same order of
magnitude as the relaxation time, is thus an important quantity:
knowing it allows to distinguish between non-relaxed systems, which
should be described by a QSS, and relaxed ones, for which collisional
effects need to be taken into account and an equilibrium description
may be relevant. This lifetime depends on the system under
consideration. Kinetic theory usually predicts a QSS lifetime of order
$N$, e.g., for 3d plasmas or 1d self-gravitating systems (see
\cite{Joyce:2010} for recent numerical tests), but this is not always
the case; for example, the QSS lifetime is of order $N/\ln N$ for 3d
self-gravitating systems \cite{Chandrasekhar:1944}. Furthermore, it is
known that the Lenard-Balescu collision term vanishes for 1d systems
which do not develop any spatial inhomogeneity (see
\cite{Eldridge:1963} for the 1d Coulomb case); one thus expects a
relaxation time much longer than $N$ in these cases.  This is indeed
numerically observed, see \cite{Yamaguchi:2004}, where a time of order
$N^{1.7}$ is reported, and \cite{Rocha:2013}, where larger systems
sizes are studied and the relaxation time is claimed to be of order
$N^2$. A similar vanishing of the Lenard-Balescu operator has been found for
  point vortices in an axisymmetric configuration \cite{Dubin:1988,Chavanis:2012a,Chavanis:2012b}.

In a recent work, QSSs have been looked for and found
  in a dynamical setting different from the ones reviewed above, 
namely, in an anisotropic Heisenberg model with mean-field
interactions \cite{Gupta:2011}. Note that similar spin models with mean field interactions have been suggested to be relevant to describe some layered spin structures \cite{Campa:2007}. Specifically, the model 
in \cite{Gupta:2011} comprises $N$
globally coupled three-component Heisenberg spins evolving under
classical spin dynamics. An associated Vlasov-type equation is
introduced, and QSSs are stationary solutions of this equation. In
addition, numerical simulations for axisymmetric QSS suggest a QSS
lifetime increasing superlinearly with $N$. In order to understand
this observation analytically, we present in this work a detailed
study of the Lenard-Balescu operator that accounts for leading
finite-$N$ corrections of order $1/N$ to the Vlasov equation. 
With respect to 1D Hamiltonian systems, the spin dynamics 
introduces a new term in the Lenard-Balescu operator; it also complicates 
the analytical structure of the dispersion relation for the Vlasov-type 
equation. Nevertheless, we can still demonstrate that corrections at order 
$1/N$ are identically zero, so
that relaxation occurs over a time longer than of order $N$, in
agreement with the reported numerical results.

The paper is structured as follows. In section \ref{model}, we describe
the model of study and its equilibrium phase diagram. As a step towards
deriving the Vlasov equation to analyze the evolution of the phase space
distribution in the limit $N \to \infty$, in section 
\ref{klimontovich}, we first write down the so-called Klimontovich equation; this leads 
to a derivation of the Vlasov equation and a discussion of a class of its
stationary solutions in section \ref{vlasov}. Section
\ref{lenard-balescu} contains our main results:
it is devoted to the derivation of the Lenard-Balescu equation for
our model of study, that is, the leading $1/N$ correction
to the Vlasov equation; we show that this correction identically 
vanishes. In
section \ref{examples}, we consider an example of a Vlasov-stationary
solution. In the energy range in which it is Vlasov stable, we
demonstrate by performing numerical simulations of the dynamics that
indeed its relaxation to equilibrium occurs over a timescale that does
not grow linearly but rather superlinearly with $N$, in support of our analysis.  The paper ends with
conclusions.
\section{The model}
\l{model}
The model studied in Ref. \cite{Gupta:2011} comprises $N$ globally coupled classical Heisenberg spins
of unit length, denoted by ${\mathbf S}_i=(S_{ix},S_{iy},S_{iz})$,
$i=1,2,\ldots,N$. In terms of spherical polar angles $\th_i \in
[0,\pi]$ and $\phi_i \in [0,2\pi]$, one has $S_{ix}=\sin \th_i \cos \ph_i, S_{iy}=\sin \th_i \sin
\ph_i, S_{iz}=\cos \th_i$. The Hamiltonian of the system is 
\be
H=-\fr{J}{2N}\sum_{i,j=1}^N {\bf S}_i \cdot {\bf
S}_j+D\sum_{i=1}^N S^2_{iz}.
\l{H}
\ee
Here, the first term with $J > 0$ describes a ferromagnetic mean-field
coupling between the spins, while the second term is the energy due to a local
anisotropy. We consider $D>0$, for which the energy is lowered by having the
magnetization 
\be
{\bf m}\equiv\fr{1}{N}\sum_{i=1}^N{\bf S}_i
\ee
pointing in the $xy$
plane. The coupling constant $J$ in equation (\ref{H}) is scaled by $N$ to make the
energy extensive \cite{Kac:1963}, but the system is non-additive, implying thereby that
it cannot be trivially subdivided into independent
macroscopic parts, as is possible with short-range systems. In this
work, we take unity for $J$ and the Boltzmann constant. 

In equilibrium, the system
(\ref{H}) shows a continuous phase transition as a function of the
energy density $e$, from a low-energy magnetized phase in which the
system is ordered in the $xy$ plane to a high-energy
non-magnetized phase, across a critical threshold given by \cite{Gupta:2011}
\be
e_c=D\Big(1-\frac{2}{\beta_c}\Big),
\l{ec}
\ee
where the inverse temperature $\beta_c$ satisfies
\be
\fr{2}{\beta_c
}=1-\fr{1}{2\beta_cD}+\fr{e^{-\beta_cD}}{\sqrt{\pi\beta_cD}
\mathrm{Erf}[\sqrt{\beta_cD}]}. \l{betac}
\ee
Here, $\mathrm{Erf}[x]=(2/\sqrt{\pi})\int_0^x {\rm d}t~ e^{-t^2}$ is the
error function.

The microcanonical dynamics of the system (\ref{H}) is given by the set of coupled first-order differential equations
\be
\fr{{\rm d}{\bf S}_i}{{\rm d}t}=\{{\bf S}_i,H\}; ~~~~i=1,2,\ldots,N.
\l{eom1}
\ee
Here, noting that the canonical variables for a
classical spin are $\phi$ and 
\be
u \equiv \cos \th,
\ee
the Poisson bracket $\{A,B\}$ for
two functions of the spins are given by
$\{A,B\}\equiv \sum_{i=1}^N (\partial A/\partial \phi_i
\partial B/\partial u_{i}-\partial A/\partial u_{i}
\partial B/\partial \phi_i)$, which may be rewritten as \cite{Mermin:1967}
\be
\{A,B\}=\sum_{i=1}^N {\bf S}_i \cdot \fr{\partial A} {\partial
{\bf S}_i}\times \fr{\partial B}{\partial {\bf S}_i}.
\l{poisson}
\ee
Using equations (\ref{eom1}) and (\ref{poisson}), we obtain the equations of
motion of the system as
\bea
&&\dot{S}_{ix}=S_{iy}m_z-S_{iz}m_y-2DS_{iy}S_{iz}, \l{eqnmotionx} \\
&&\dot{S}_{iy}=S_{iz}m_x-S_{ix}m_z+2DS_{ix}S_{iz}, \l{eqnmotiony}\\
&&\dot{S}_{iz}=S_{ix}m_y-S_{iy}m_x, \l{eqnmotionz}
\eea
where the dots denote derivative with respect to time.
Summing over $i$ in equation (\ref{eqnmotionz}), we find that
$m_z$ is a constant of motion. The dynamics also conserves the total
energy and the length of each spin. Using equations (\ref{eqnmotionx}),
(\ref{eqnmotiony}), and (\ref{eqnmotionz}), we obtain the time evolution of
the variables $\th_i$ and $\ph_i$ as
\bea
\dot{\th_i}&=&m_x \sin \ph_i-m_y \cos \ph_i,\l{eqnmotiontheta} \\ 
\dot{\ph_i}&=&m_x\cot \th_i \cos \ph_i+m_y \cot \th_i \sin \ph_i
-m_z+2D\cos \th_i. \l{eqnmotionphi}
\eea
\section{The Klimontovich equation}
\l{klimontovich}
The state of the $N$-spin system is described by the discrete one-spin time-dependent density function
\be
f_d(u,\ph,t)=\fr{1}{N}\sum_{i=1}^N\delta(u-u_i(t))\delta(\ph-\ph_i(t)),
\l{fd}
\ee
which is defined such that $f_d(u,\ph,t){\rm d}u {\rm d}\ph$ counts the
number of spins with its canonical coordinates in $[u,u+du]$ and $[\ph, \ph+d\ph]$. Here, $\delta$ is the Dirac delta function,
$(u,\ph)$ are the Eulerian coordinates of the phase space, while
$(u_i,\ph_i)$ are the Lagrangian coordinates of the spins. Note that
$f_d$ satisfies $f_d(u,\ph,t)=f_d(u,\ph+2\pi,t)$ and the normalization
$\int_0^{2\pi} {\rm d}\ph \int_{-1}^1 {\rm d}u ~f_d(u,\ph,t)=1$.

Differentiating $f_d$ with respect to time and using the equations of
motion, (\ref{eqnmotiontheta}) and (\ref{eqnmotionphi}), we get the Klimontovich equation
for the time evolution of $f_d$ as
\bea
&&\fr{\partial f_d(u,\ph,t)}{\partial t}=-g_u\fr{\partial }{\partial u}f_d(u,\ph,t)-g_\phi\fr{\partial }{\partial
\phi}f_d(u,\ph,t), 
\l{Klimontovich}
\eea
where 
\bea
&&g_u \equiv g_u[f_d](u,\phi)\nonumber \\
&&=\sqrt{1-u^2}(m_y[f_d]\cos \ph-m_x[f_d]\sin \ph),
\l{gu} \\
&&g_\phi \equiv g_\phi[f_d](u,\phi)\nonumber \\
&&=m_x[f_d]\fr{u}{\sqrt{1-u^2}}\cos \ph+m_y[f_d]\fr{u}{\sqrt{1-u^2}}\sin \ph-m_z[f_d]+2Du, \l{gphi} \\
&&(m_x,m_y,m_z)[f_d]\nonumber \\
&&=\int_0^{2\pi}{\rm d}\phi \int_{-1}^1 {\rm d}u~(\sqrt{1-u^2}\cos \ph,\sqrt{1-u^2} \sin
\ph,u)f_d(u,\ph,t) \l{mxmymz}.
\eea
\section{The Vlasov equation, and a class of stationary solutions}
\l{vlasov}
We now define an averaged one-spin density function $f_0(u,\phi,t)$,
corresponding to averaging $f_d(u,\phi,t)$ over an ensemble of initial conditions
close to the same macroscopic initial state. We write, quite generally, for an initial condition of the ensemble that
\be
f_d(u,\phi,t)=f_0(u,\phi,t)+\fr{1}{\sqrt{N}}\delta f(u,\phi,t),
\l{expansion}
\ee
where, denoting by angular brackets the averaging with respect to the
initial ensemble, we have $\langle f_d\rangle=f_0$. Here, $\delta f$
gives the difference between $f_d$, which depends on the given initial
condition, and $f_0$, which depends on the average with respect to
the ensemble of initial conditions.

Using equation (\ref{expansion}) in equation (\ref{Klimontovich}), we get
\bea
&&\fr{\partial f_0}{\partial t}+g_u^0\fr{\partial f_0}{\partial
u}+g_\phi^0\fr{\partial f_0}{\partial \phi}+\fr{1}{N}\Big[ \delta
g_u \fr{\partial \delta f}{\partial u}+\delta g_\phi \fr{\partial \delta
f}{\partial \phi}\Big]\nonumber \\
&&+\fr{1}{\sqrt{N}}\Big[\fr{\partial \delta f}{\partial t}+\delta g_u \fr{\partial
f_0}{\partial u}+g_u^0 \fr{\partial \delta f}{\partial u}+\delta g_\phi \fr{\partial
f_0}{\partial \phi}+g_\phi^0 \fr{\partial \delta f}{\partial
\phi}\Big]=0, 
\l{before-Vlasov0}
\eea
where $g_u^0=g_u[f_0], g_\phi^0=g_\phi[f_0]$, and
\bea
&&\delta g_u=\sqrt{1-u^2}(m_y[\delta f]\cos
\ph-m_x[\delta f]\sin \ph), \l{deltagu} \\
&&\delta g_\phi=m_x[\delta f]\fr{u}{\sqrt{1-u^2}}\cos
\ph+m_y[\delta f]\fr{u}{\sqrt{1-u^2}}\sin \ph-m_z[\delta
f].\l{deltagphi}
\eea
We now average equation (\ref{before-Vlasov0}) with respect to the ensemble of initial conditions, and note that $\langle\delta f\rangle=0$ implies $\langle m_x[\delta f] \rangle = \langle m_y[\delta f] \rangle = \langle m_z[\delta f] \rangle=0$. Thus $\langle \delta g_u \rangle=\langle \delta g_\phi \rangle=0$, and we get
\bea
&&\fr{\partial f_0}{\partial t}+g_u^0\fr{\partial f_0}{\partial
u}+g_\phi^0\fr{\partial f_0}{\partial \phi}=-\fr{1}{N}\Big\langle \delta
g_u \fr{\partial \delta f}{\partial u}+\delta g_\phi \fr{\partial \delta
f}{\partial \phi}\Big\rangle. 
\l{before-Vlasov}
\eea
For finite times and in the limit $N \to \infty$ (or, for times $t \ll N$), we obtain the Vlasov equation satisfied by the averaged
one-spin density function $f_0$ \cite{Nicholson:1992} as
\bea
&&\fr{\partial f_0}{\partial t}+g_u^0\fr{\partial f_0}{\partial
u}+g_\phi^0\fr{\partial f_0}{\partial \phi}=0. 
\l{Vlasov}
\eea

Note that the Vlasov equation has been formally obtained after averaging
over an ensemble of initial conditions. However, if the fluctuations in the initial
conditions are weak and do not grow too fast in time, we expect the
Vlasov equation to also describe the time evolution of a {\it single}
 initial condition in the limit $N\to \infty$. This
is put on firm mathematical grounds in
\cite{Neunzert:1972,Braun:1977,Dobrushin:1979}, for systems with a
standard kinetic energy and a regular enough interaction potential.

From equations (\ref{gu}), (\ref{gphi}), and (\ref{mxmymz}), it is clear that any distribution
 that does not depend on the angle $\phi$ (thus, axisymmetric about the
$z$-axis) is a stationary solution of the Vlasov
equation~(\ref{Vlasov}).

\section{The Lenard-Balescu equation}
\l{lenard-balescu}

We obtain in this section the main result of our paper: for model (\ref{H}), the Lenard-Balescu operator, computed for stationary solutions of the Vlasov equation of the form $f_0(u)$, identically vanishes.

\subsection{Formal derivation}
\l{derivation}
The Lenard-Balescu equation describes the slow evolution of a stable stationary solution of the Vlasov equation under the influence of
finite-$N$ corrections, at leading order in $1/N$ \cite{Nicholson:1992}. From
equation (\ref{before-Vlasov}), we get the Lenard-Balescu equation as  
\bea
&&\fr{\partial f_0}{\partial t}=-\fr{1}{N}\Big\langle \delta
g_u \fr{\partial \delta f}{\partial u}+\delta g_\phi \fr{\partial \delta
f}{\partial \phi}\Big\rangle, 
\l{LB}
\eea
where, subtracting equation (\ref{Vlasov}) from equation
(\ref{before-Vlasov0}), and keeping only the terms of order 
$1/\sqrt{N}$, we find that $\delta f$ follows the Vlasov
equation linearized around its stable stationary solution $f_0$:
\bea
\fr{\partial \delta f}{\partial t}=\sqrt{1-u^2}\Big[\delta m_x\sin
\ph-\delta m_y\cos \ph\Big] \fr{\partial
f_0}{\partial u}- (2Du-m_z[f_0]) \fr{\partial \delta f}{\partial \phi}. 
\l{linear-Vlasov}
\eea
Here, we have used $g_\phi^0=2Du-m_z[f_0]$ and $\delta m_{x,y}\equiv
m_{x,y}[\delta f]$.  Now, a natural timescale separation hypothesis
greatly reduces the complexity of finding the solutions of the coupled
system of PDEs, equations (\ref{LB})
and (\ref{linear-Vlasov}): The first of the two equations evolves on a slow
$O(1/N)$ timescale, while the second one evolves on a fast $O(1)$ timescale.
Then, we may first solve equation (\ref{linear-Vlasov}), and then use
its solution to compute the right-hand side of (\ref{LB}) in the limit $t
\to \infty$. 

At this point, it is useful to make a comparison of our case with the standard
case of particles with a kinetic energy moving in a classical
potential, for example, a 1d system of particles with Coulomb
interactions. The equivalent of the axisymmetric stationary solutions (independent of
$\phi$) introduced in section ~\ref{vlasov} are the
homogeneous solutions which depend only on velocity in this standard
setting, so that the analog of $\delta
g_{\phi}$ vanishes, whereas in our case of the spin dynamics, we have to
deal with the extra term $\Big\langle \delta g_\phi \fr{\partial
  \delta f}{\partial \phi}\Big\rangle$.

\subsection{Solution of the linearized Vlasov equation}
\l{linearizedvlasov}
We now solve equation (\ref{linear-Vlasov}) for $\delta f$, using Fourier-Laplace
transforms
\bea
&&\delta f(u,\phi,t)=\sum_{k=-\infty}^\infty\int_\Gamma
\fr{{\rm d}\omega}{2\pi} \widetilde{\delta f}_k(u,\omega)e^{i(k\phi-\omega
t)}, \l{deltafuphi-FT} \\
&&\widetilde{\delta f}_k(u,\omega)=\int_0^{2\pi}\fr{{\rm d}\phi}{2\pi}
\int_0^\infty {\rm d}t~\delta f(u,\phi,t)e^{-i(k\phi-\omega
t)},
\eea
where the Laplace contour $\Gamma$ is a horizontal line in the
complex-$\omega$ plane that passes above all singularities of
$\widetilde{\delta f}_k(u,\omega)$. 

We have
\bea
&&\delta m_x=\int_\Gamma \fr{{\rm d}\omega}{2\pi}e^{-i\omega
t}\int_{-1}^1 {\rm d}u ~\pi \sqrt{1-u^2}\Big(\widetilde{\delta
f}_{-1}(u,\omega)+\widetilde{\delta
f}_{+1}(u,\omega)\Big), \l{deltamx} \\
&&\delta m_y=\int_\Gamma \fr{{\rm d}\omega}{2\pi}e^{-i\omega
t}\int_{-1}^1 {\rm d}u ~\fr{\pi}{i} \sqrt{1-u^2}\Big(\widetilde{\delta
f}_{-1}(u,\omega)-\widetilde{\delta
f}_{+1}(u,\omega)\Big),
\l{deltamy} \\
&&\delta m_z=m_z[\delta f]=\int_\Gamma \fr{{\rm d}\omega}{2\pi}e^{-i\omega
t}\int_{-1}^1 {\rm d}u ~\pi u\widetilde{\delta
f}_0(u,\omega),
\l{deltamz}
\eea
so that equation (\ref{linear-Vlasov}) gives
\bea
&&\widetilde{\delta f}_{\pm 1}(u,\omega)=-\fr{\pi
\sqrt{1-u^2}f_0'(u)}{2Du-m_z[f_0]\mp\omega}\int_{-1}^1 {\rm d}u'
\sqrt{1-u'^2}~\widetilde{\delta f}_{\pm 1}(u',\omega)\nonumber \\
&&\mp\fr{i\delta
f_{\pm 1}(u,0)}{2Du-m_z[f_0] \mp
\omega},
\l{deltaf0}
\eea
where $f_0'(u)=\partial f_0(u)/\partial u$ and $\delta f_{\pm 1}(u,0)$
is the Fourier transform of the initial fluctuations $\delta
f(u,\phi,0)$. Multiplying both sides of the above equation by $\sqrt{1-u^2}$ and then integrating over $u$,
we get
\be
\eps_{\pm
1}(\omega)\int_{-1}^1 {\rm d}u \sqrt{1-u^2}~\widetilde{\delta f}_{\pm
1}(u,\omega)=\mp i \int_{-1}^1 {\rm d}u
\fr{\sqrt{1-u^2}\delta f_{\pm 1}(u,0)}{2Du-m_z[f_0] \mp \omega},
\l{deltaf-int}
\ee
where $\eps_{\pm 1}(\omega)$ is the so-called ``Plasma response dielectric function"
\cite{Nicholson:1992}:
\be
\eps_{\pm 1}(\omega)=1+\pi \int_{LC} {\rm d}u \fr{(1-u^2)f_0'(u)}{2Du
-m_z[f_0] \mp \omega}.
\l{eps-defn}
\ee
To make the dielectric function $\eps_{+1}$ (also, $\eps_{-1}$) analytic 
in the vicinity of the real axis (${\rm Im}(\omega)=0$), which will be 
needed for later purpose, the above integral has to be performed along the Landau 
contour shown in Fig. \ref{LC}, as discussed in \cite{Nicholson:1992}; we have in 
this case
\bea
\eps_{\pm 1}(\omega)=\left\{ 
\begin{array}{l}
                1+\pi \int_{LC} {\rm d}u
                \fr{(1-u^2)f_0'(u)}{2Du-m_z[f_0]\mp\omega};({\rm Im}(\omega) >0), \\ \\
                1+\pi {\rm P} \int_{LC} {\rm d}u
                \fr{(1-u^2)f_0'(u)}{2Du-m_z[f_0]\mp\omega}\nonumber \\ \\
                \pm i\fr{\pi^2}{2D}
                \left.(1-u^2)f_0'(u)\right|_{(m_z[f_0]\pm\omega)/(2D)}; ({\rm Im}(\omega) =0), \nonumber \\ \\
                1+\pi \int_{LC} {\rm d}u
                \fr{(1-u^2)f_0'(u)}{2Du-m_z[f_0]\mp\omega}\nonumber \\ \\
                \pm i\fr{2\pi^2}{2D}
                \left.(1-u^2)f_0'(u)\right|_{(m_z[f_0]\pm\omega)/(2D)};({\rm Im}(\omega) < 0), \nonumber \\
               \end{array}
        \right. \\
\l{eps-explicit}
\eea
where ${\rm P}$ denotes the principal part.  If $\omega \notin
[-2D-m_z[f_0]; 2D-m_z[f_0]]$ (respectively,  $\omega \notin [-2D+m_z[f_0];
2D+m_z[f_0]]$), equation (\ref{eps-defn}) already defines an analytic
function $\eps_{+1}$ (respectively. $\eps_{-1}$) in the vicinity of a real
$\omega$, without the need to take into account extra pole
contributions as in (\ref{eps-explicit}).
Note that $\eps_{\pm 1}(\omega)$ has 
two branch cut singularities on the real axis at $\omega=2D-m_z[f_0]$ and
$\omega=-2D-m_z[f_0]$, which is the reason why these functions may be seen as 
multi-valued in the lower-half $\omega$-plane.

\begin{figure}[here!]
\centering
\includegraphics[width=100mm]{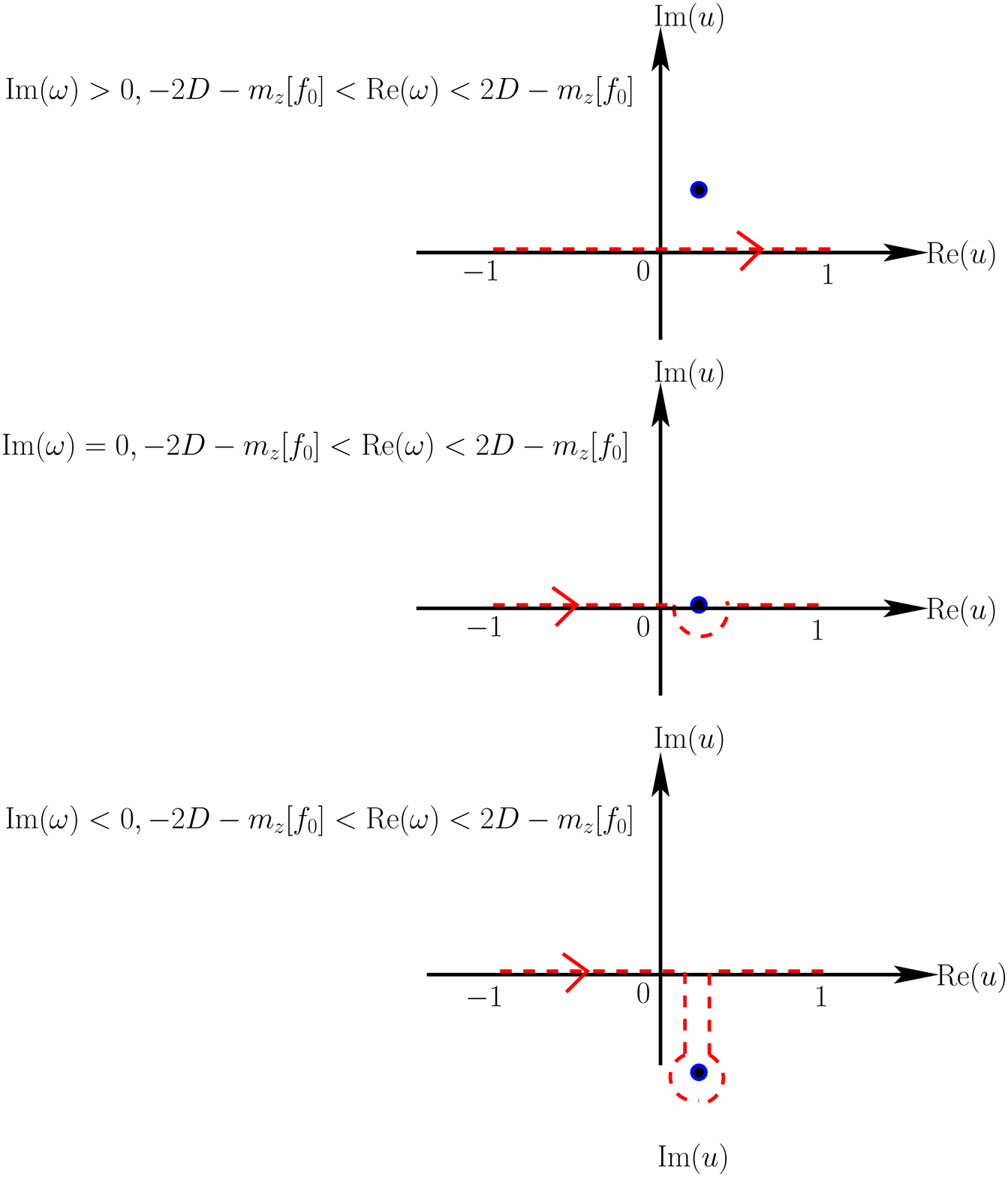}
\caption{The Landau contour LC to evaluate $\eps_{+1}(\omega)$, shown as the dashed red line in the
complex-$u$ plane.}
\l{LC}
\end{figure}

Using equation (\ref{deltaf-int}) in equation (\ref{deltaf0}) gives
\bea
&&\widetilde{\delta f}_{\pm 1}(u,\omega)=\pm \fr{i\pi
\sqrt{1-u^2}f_0'(u)}{\eps_{\pm
}(\omega)[2Du-m_z[f_0]\mp\omega]}\int_{-1}^1 {\rm d}u
\fr{\sqrt{1-u^2}\delta f_{\pm 1}(u,\phi,0)}{2Du -m_z[f_0]\mp
\omega}\nonumber \\
&&\mp\fr{i\delta f_{\pm 1}(u,0)}{2Du-m_z[f_0] \mp
\omega}.
\l{deltaf}
\eea
We see from the above expression that the real pole at $\omega=\pm
(2Du-m_z[f_0])$ is due to the
free part of the evolution that does not involve interaction among the
spins, and results in undamped oscillations of the fluctuations
$\delta f(u,\phi,t)$, see equation (\ref{deltafuphi-FT}).
The other set of poles corresponds to the zeros of the dielectric
function $\eps_{\pm }(\omega)$, i.e., values 
$\omega_{\rm p}$ (complex in general) that satisfy
\be
\eps_{\pm 1}(\omega_{\rm p})=0.
\l{dispersion-relation-1}
\ee
Equation (\ref{deltafuphi-FT}) implies that these poles determine the
growth or decay of the fluctuations $\delta f(u,\phi,t)$ in time, depending on
their location in the complex-$\omega$ plane. For example,
when the poles lie in the upper-half complex $\omega$-plane, the
fluctuations grow in time. On the other hand, when the poles are either on or below the real-$\omega$ axis,
the fluctuations do not grow in time, but rather oscillate or decay in
time, respectively. Then, the condition ensuring linear stability of a 
stationary solution of the Vlasov equation reads   
\be
\eps_{\pm 1}(\omega_{\rm p})=0 ~\Rightarrow~ {\rm Im}(\omega_{\rm p}) \le 0.
\l{dispersion-relation}
\ee
The condition ${\rm Im}(\omega_{\rm p}) = 0$ corresponds to marginal
stability. 

To end this section, let us define for later use the
quantities
\be
\delta m_\pm \equiv \delta m_x\pm i\delta m_y.
\ee
On using equation (\ref{deltaf-int}), we get the corresponding Laplace transforms as
\bea
&&\widetilde{\delta m}_+(\omega)=
\fr{2i\pi}{\eps_{-1}(\omega)}\int_{-1}^1 {\rm d}u
\fr{\sqrt{1-u^2}\delta f_{- 1}(u,0)}{2Du-m_z[f_0] + \omega}, \l{deltam+} \\
&&\widetilde{\delta m}_-(\omega)=
-\fr{2i\pi}{\eps_{+1}(\omega)}\int_{-1}^1 {\rm d}u
\fr{\sqrt{1-u^2}\delta f_{ +1}(u,0)}{2Du-m_z[f_0] - \omega}. \l{deltam-}
\eea
Equation (\ref{deltaf}) may now be expressed in terms of
$\widetilde{\delta m}_\pm$ as
\bea
\widetilde{\delta f}_{+ 1}(u,\omega)=-
\fr{\sqrt{1-u^2}f_0'(u)}{2[2Du-m_z[f_0]-\omega]}\widetilde{\delta m}_-(\omega)-\fr{i\delta
f_{+1}(u,0)}{2Du-m_z[f_0] -
\omega}, \l{deltafp1}\\
\widetilde{\delta f}_{- 1}(u,\omega)=-
\fr{\sqrt{1-u^2}f_0'(u)}{2[2Du-m_z[f_0]+\omega]}\widetilde{\delta m}_+(\omega)+\fr{i\delta
f_{-1}(u,0)}{2Du-m_z[f_0] +
\omega}. 
\l{deltafm1}
\eea

\subsection{Computing the Lenard-Balescu operator}
\l{lenardbalescu-computation}
We now compute the Lenard-Balescu operator, given by the right hand
side of equation (\ref{LB}), in the limit $t \to \infty$, by using the results of the
preceding subsection. We have
\bea
\Big\langle \delta
g_\phi \fr{\partial \delta f}{\partial
\phi}\Big\rangle=\sum_{k,l}\int_\Gamma \int_{\Gamma'}
\fr{{\rm d}\omega {\rm d}\omega'}{4\pi^2}e^{i(k+l)\phi}e^{-i(\omega+\omega')t}il\Big\langle \widetilde{\delta
g}_{\phi,k}(u,\omega) \widetilde{\delta
f}_l(u,\omega') \Big\rangle,\l{delgphidelf}\\
\Big\langle \delta
g_u \fr{\partial \delta f}{\partial u}\Big\rangle=\sum_{k,l}\int_\Gamma
\int_{\Gamma'}
\fr{{\rm d}\omega {\rm d}\omega'}{4\pi^2}e^{i(k+l)\phi}e^{-i(\omega+\omega')t}\Big\langle \widetilde{\delta
g}_{u,k}(u,\omega)\fr{\partial \widetilde{\delta
f}_l(u,\omega')}{\partial u}\Big\rangle. \l{delgudelf}
\eea
From equations (\ref{deltagu}) and (\ref{deltagphi}), we have
\bea
&&\delta g_\phi=\Big(\fr{\delta m_++\delta m_-}{2}\Big)\cos
\ph\fr{u}{\sqrt{1-u^2}}+\Big(\fr{\delta m_+-\delta m_-}{2i}\Big)\sin
\ph\fr{u}{\sqrt{1-u^2}}-\delta m_z[\delta f]\l{deltagphi0}, \nonumber \\ \\
&&\delta g_u =\sqrt{1-u^2}\Big[\Big(\fr{\delta m_+-\delta m_-}{2i}\Big)\cos
\ph-\Big(\fr{\delta m_++\delta m_-}{2}\Big)\sin \ph\Big], \l{deltagu0} 
\eea
so that we have
\bea
&&\widetilde{\delta
g}_{\phi,\pm 1}(u,\omega)=\widetilde{\delta
m}_\mp (\omega)\fr{u}{2\sqrt{1-u^2}},\widetilde{\delta
g}_{\phi,0}(u,\omega)=\widetilde{\delta
m}_z(\omega), \l{deltagphi-FT} \\
&&\widetilde{\delta
g}_{u,\pm 1}(u,\omega)=\mp\fr{\sqrt{1-u^2}}{2i}\widetilde{\delta
m}_\mp(\omega). \l{deltagu-FT} 
\eea

From equations (\ref{deltam+}) and (\ref{deltam-}), we see that
$\widetilde{\delta m}_+$ (respectively, $\widetilde{\delta m}_-$)
depends on $\delta f_{- 1}(u,0)$ (respectively, $\delta f_{+1}(u,0)$). Then, to
compute the right hand hand side of equations (\ref{delgphidelf}) and
(\ref{delgudelf}), we need to evaluate averages of the type $\langle \delta f_k(u,0) \delta f_l(v,0)\rangle$ for the initial fluctuations at $t=0$. Note
that $t=0$ as a notation is somewhat inappropriate, since actually this computation has to be repeated for any value of the slow time. We may assume that at $t=0$, the
spins are almost independent (i.e., the two-spin correlation is of order $1/N$). In
this case, one gets
\be
\langle \delta f_k(u,0) \delta f_l(v,0)\rangle = \frac{\delta_{k,-l}}{2\pi}\left[f_0(u)\delta(u-v)
+h(u,v)\right],
\l{autocorrel}
\ee
where $\delta_{k,-l}$ is the Kronecker delta function, and $h(u,v)$ is a
smooth function. The precise form of this undetermined smooth function
will play no role in the computation, as we will show below.

\subsubsection{Computing $\Big\langle \delta
g_\phi \fr{\partial \delta f}{\partial
\phi}\Big\rangle$}
Combining equations (\ref{deltagphi-FT}) and (\ref{autocorrel}), we see that
on the right hand side of equation (\ref{delgphidelf}), only the terms
$(k=+1,l=-1)$ and $(k=-1,l=+1)$ give a non-zero contribution. We detail
below the computation for the $(k=+1,l=-1)$ case, the other being similar. 
Note that $\Big\langle \delta g_\phi \fr{\partial \delta f}{\partial \phi}\Big\rangle$ is real; thus, we have to compute only the real part of 
the $(k=+1,l=-1)$ term, since its imaginary part must cancel with that of the $(k=-1,l=+1)$ term.
In the following computation, we set $h(u,v)=0$; we will check at the end that indeed the contributions containing $h$ vanish.

We have
\begin{eqnarray}
&&\langle \widetilde{\delta g}_{\phi,+1}(u,\omega) \widetilde{\delta
f}_{-1}(u,\omega')\rangle= -\frac{uf'_0(u)}{4(2Du-m_z[f_0]+\omega')}\langle \widetilde{\delta m}_-(\omega) 
\widetilde{\delta m}_+(\omega')\rangle \nonumber \\
&&+i\frac{u}{2\sqrt{1-u^2}(2Du-m_z[f_0]+\omega')} \langle \widetilde{\delta m}_-(\omega) 
\delta f_{-1}(u,0) \rangle.  
\l{LB1a}
\end{eqnarray}
We thus need 
\begin{eqnarray}
&&\langle \widetilde{\delta m}_-(\omega) \widetilde{\delta
m}_+(\omega')\rangle \nonumber \\
&&= \frac{2\pi}{\epsilon_{+1}(\omega)\epsilon_{-1}(\omega')}
\int_{-1}^1 {\rm d}v
\frac{(1-v^2)f_0(v)}{(2Dv-m_z[f_0]-\omega)(2Dv-m_z[f_0]+\omega')},
\l{corr-1} \\
&&\langle \widetilde{\delta m}_-(\omega) \delta f_{-1}(u,0)\rangle =
-i\frac{\sqrt{1-u^2}f_0(u)}{\epsilon_{+1}(\omega)(2Du-m_z[f_0]-\omega)},
\l{corr-2}
\end{eqnarray}
where we have used equation (\ref{autocorrel}). Using these equations in
equation (\ref{LB1a}), we obtain
\begin{eqnarray}
&&\langle \widetilde{\delta g}_{\phi,+1}(u,\omega) \widetilde{\delta
f}_{-1}(u,\omega')\rangle\nonumber \\
&&= -\frac{\pi
uf'_0(u)}{2(2Du-m_z[f_0]+\omega')\epsilon_{+1}(\omega)\epsilon_{-1}(\omega')}\nonumber
\\
&&\times  \int_{-1}^1
{\rm d}v
\frac{(1-v^2)f_0(v)}{(2Dv-m_z[f_0]-\omega)(2Dv-m_z[f_0]+\omega')} \nonumber \\
&&+\frac{uf_0(u)}{2(2Du-m_z[f_0]+\omega')(2Du-m_z[f_0]-\omega)\epsilon_{+1}(\omega)}
\nonumber \\
&&\equiv b_1+b_2.
\end{eqnarray}

Now, to compute the contribution to $\Big\langle \delta
g_\phi \fr{\partial \delta f}{\partial \phi}\Big\rangle$ for
($k=+1,l=-1$), we have to integrate $b_1$ and $b_2$ over $\omega$
and $\omega'$; we define
\begin{eqnarray}
B_1 \equiv \int_\Gamma \int_{\Gamma'}\frac{{\rm d}\omega
{\rm d}\omega'}{4\pi^2}(-i)e^{-i(\omega+\omega')t}b_1, \l{B1}
B_2 \equiv \int_\Gamma \int_{\Gamma'}\frac{{\rm d}\omega
{\rm d}\omega'}{4\pi^2}(-i)e^{-i(\omega+\omega')t}b_2. \l{B2}
\end{eqnarray}
We want to compute $B_1$ and $B_2$ in the limit $t\to \infty$; thus, we
will discard all terms decaying for large $t$. 

Let us start with $B_1$. We have
\bea
&&B_1=i\frac{\pi u
f_0'(u)}{2}\int_{\Gamma'}\frac{{\rm d}\omega'}{2\pi}\frac{e^{-i\omega'
t}}{(2Du-m_z[f_0]+\omega')(2Dv-m_z[f_0]+\omega')\epsilon_{-1}(\omega')}\nonumber \\
&&\times \int_{\Gamma}\frac{{\rm d}\omega}{2\pi}e^{-i\omega
t}\int_{-1}^1 {\rm
d}v~
\frac{(1-v^2)f_0(v)}{\eps_{+1}(\omega)(2Dv-m_z[f_0]-\omega)}.
\l{B1-1}
\eea
\begin{figure}[here!]
\centering
\includegraphics[width=100mm]{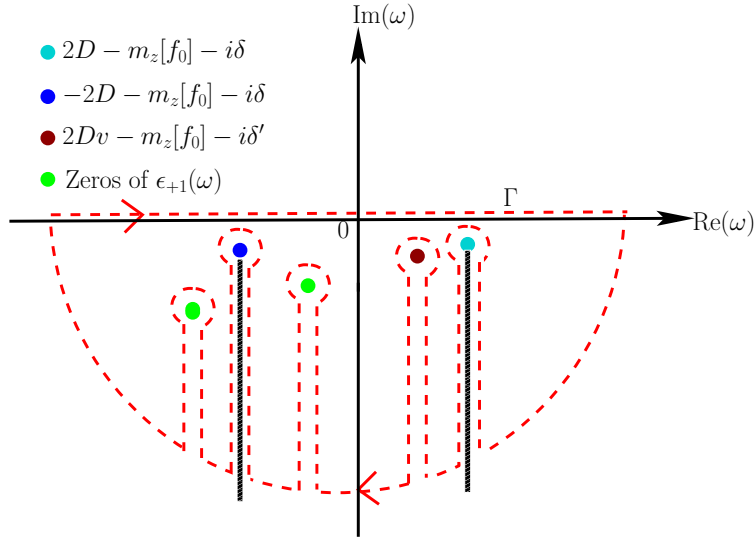}
\caption{The $\Gamma$ contour to evaluate the integral in equation (\ref{integ_omega1}). Here, $\delta \to 0$ and $\delta' \to 0$ are infinitesimal positive
numbers. The thick vertical lines denote branch-cut singularities.}
\l{omega}
\end{figure}

First, note that for large $t$,
\be
\int_\Gamma \frac{{\rm d}\omega}{2\pi}\frac{e^{-i\omega
    t}}{\epsilon_{+1}(\omega)(2Dv-m_z[f_0]-\omega)} \sim
    i\frac{e^{-i(2Dv-m_z[f_0])t}}
{\epsilon_{+1}(2Dv-m_z[f_0])},
\l{integ_omega1}
\ee
which is obtained by deforming the $\Gamma$ contour into the ${\rm
  Im}(\omega)<0$ half-plane, see Fig. \ref{omega}, and noting that due
to the $e^{-i\omega t}$ factor, the only contribution to the contour
integral in the limit $t \to \infty$ comes from the pole
$2Dv-m_z[f_0]-i\delta'$. Indeed, since $f_0$ is assumed stable, the
zeros of $\epsilon_{+1}$, if any, have a negative imaginary part;
their contributions thus decay exponentially in time. Similarly, the
branch-cut singularities, see
Fig.~\ref{omega}, contribute terms decaying algebraically in time. Note
also that the important singularity at $2Dv-m_z[f_0]$ is
real, and is in the range $[-2D-m_z[f_0]; 2D-m_z[f_0]]$. Then, according to the
discussion in section~\ref{linearizedvlasov}, one needs to use the expression
(\ref{eps-explicit}) for $\epsilon_{+1}$. The same remark applies to all the
computations below, and we will not recall it each time.

The integration over $\omega'$ in equation (\ref{B1-1}) is a bit more complicated, since there are 
two poles on the real axis, at $\omega'=-2Du+m_z[f_0]$ and $\omega'=-2Dv+m_z[f_0]$. With the same method as
for $\omega$, we obtain
\bea
&&\int_{\Gamma'} \frac{{\rm d}\omega'}{2\pi}\frac{e^{-i\omega'
    t}}{(2Du-m_z[f_0]+\omega')(2Dv-m_z[f_0]+\omega')\epsilon_{-1}(\omega')} \nonumber \\
    &&=
-i\left[\frac{e^{i(2Du-m_z[f_0])t}}{2D(v-u)\epsilon_{-1}(-2Du+m_z[f_0])}-\frac{e^{i(2Dv-m_z[f_0])t}}{2D(v-u)
\epsilon_{-1}(-2Dv+m_z[f_0])}\right].\l{integ_omega2}
\eea
We finally get
\bea
&&B_1 = i\frac{\pi u f_0'(u)}{2}\int_{-1}^1 {\rm d}v~
\frac{(1-v^2)f_0(v)}{\epsilon_{+1}(2Dv-m_z[f_0])}\nonumber \\
&&\times\left\{\frac{e^{-i2D(v-u)t}}{2D(v-u)\epsilon_{-1}(-2Du+m_z[f_0])}-
\frac{1}{2D(v-u)\epsilon_{-1}(-2Dv+m_z[f_0])} \right\}. 
\eea
To perform the integral over $v$, we will use the following lemma.

\medskip
{\bf Lemma:}\\
\be
\lim_{t\to\infty}\int_{-1}^1 {\rm d}v \frac{\varphi(v)}{v-u}\left(
\frac{e^{-i2D(v-u)t}}{\epsilon_{-1}(-2Du+m_0)}-\frac{1}{\epsilon_{-1}(-2Dv+m_0)}
\right)=-i \pi\frac{\varphi(u)}{\epsilon_{-1}(-2Du+m_0)}.
\l{lemma}
\ee
{\it Proof:}
\bea
&&\int_{-1}^{1} {\rm d}v \frac{\varphi(v)}{v-u}\left(
\frac{e^{-i2D(v-u)t}}{\epsilon_{-1}(-2Du+m_0)}-\frac{1}{\epsilon_{-1}(-2Dv+m_0)}
\right) \nonumber \\
&&= \int_{-2D(1+u)t}^{2D(1-u)t}{\rm d}x\frac{\varphi(u+\frac{x}{2Dt})}{x}\left( 
\frac{e^{-ix}}{\epsilon_{-1}(-2Du+m_0)}-\frac{1}{\epsilon_{-1}(-2Du+m_0-\frac{x}{t})} 
\right).
\eea
Recalling that $-1<u<1$, and taking the limit $t \to \infty$, the above
expression simplifies to 
\bea
&&\int_{-\infty}^{\infty}{\rm d}x\frac{\varphi(u)}{x}\left(
  \frac{e^{-ix}}{\epsilon_{-1}(-2Du+m_0)}-\frac{1}{\epsilon_{-1}(-2Du+m_0)}\right)\nonumber
  \\
&&=
\frac{\varphi(u)}{\epsilon_{-1}(-2Du+m_0)}\int_{-\infty}^{\infty}{\rm d}x
\frac{e^{-ix}-1}{x}.
\eea
The last integral is $-i\pi$, which completes the proof of the lemma.
\medskip

Using the lemma, we conclude that
\be
B_1
=\frac{\pi^2 uf_0'(u)(1-u^2)f_0(u)}{4D\epsilon_{+1}(2Du-m_z[f_0])
\epsilon_{-1}(-2Du+m_z[f_0])}. 
\l{B1-final}
\ee

We now turn to the computation of $B_2$. The integration over $\omega$ and 
$\omega'$ is performed as above, deforming the contours in the lower-half 
$\omega'$-plane, and keeping only the contributions of the poles on the real axis.
We obtain
\bea
B_2 &=& -i\frac{uf_0(u)}{2}\frac{1}{\epsilon_{+1}(2Du-m_z[f_0])}\nonumber
\\
&=&-i\frac{uf_0(u)}{2}\frac{\epsilon_{-1}(-2Du+m_z[f_0])}{\epsilon_{+1}(2Du-m_z[f_0])\epsilon_{-1}(-2Du+m_z[f_0])}.
\eea
As explained above, we need to compute only the real part of $B_2$; thus, we keep only the contribution
coming from the imaginary part of
$\epsilon_{-1}(-2Du+m_z[f_0])$.
Using from equation (\ref{eps-explicit}) that
\be
{\rm Im}[\epsilon_{-1}(-2Du+m_z[f_0])] = - \frac{\pi^2}{2D}(1-u^2)f'_0(u),
\l{eps-minus1}
\ee
we conclude that
\be
{\rm Re}(B_2) = -\frac{\pi^2
uf_0(u)(1-u^2)f'_0(u)}{4D\epsilon_{+1}(2Du-m_z[f_0])\epsilon_{-1}(-2Du+m_z[f_0])}.
\l{B2-final}
\ee
From equations (\ref{B1-final}) and (\ref{B2-final}), we find that
$B_1+{\rm Re}(B_2)=0$. Similar to above, one can show that the real part of the contribution to $\Big\langle \delta
g_\phi \fr{\partial \delta f}{\partial \phi}\Big\rangle$ from
($k=-1,l=+1$) also vanishes, while, as discussed above, the imaginary
part of the contribution to $\Big\langle \delta
g_\phi \fr{\partial \delta f}{\partial \phi}\Big\rangle$ from
($k=-1,l=+1$) must cancel that from ($k=+1,l=-1$). So, we conclude that
for our model,
\be
\Big\langle \delta
g_\phi \fr{\partial \delta f}{\partial \phi}\Big\rangle=0.
\l{LB1}
\ee

We need now to check that the contributions containing the function $h$ introduced in (\ref{autocorrel}) indeed vanish.
For example, let us compute its contribution to $B_2$. First, its contribution to
$\langle \widetilde{\delta m}_-(\omega)\delta f_{-1}(u,0)\rangle$ is
\be 
-\frac{i}{\epsilon_{+1}(\omega)}\int_{-1}^1{\rm
d}v\frac{\sqrt{1-v^2}}{2Dv-m_z[f_0]-\omega}h(v,u).
\ee
Thus, its contribution to $B_2$ is
\begin{eqnarray}
\frac{u}{2\sqrt{1-u^2}}\int_{-1}^1 {\rm d}v \sqrt{1-v^2}h(v,u)\int_\Gamma \int_{\Gamma'}\frac{{\rm d}\omega
{\rm
d}\omega'}{4\pi^2}e^{-i(\omega+\omega')t}\frac{1}{\epsilon_{+1}(\omega)(2Dv-m_z[f_0]-\omega)}&& \nonumber\\
\times \frac{1}{(2Du-m_z[f_0]+\omega')}. &&
\end{eqnarray}
The integrals over $\omega$ and $\omega'$ can be performed as before. Since the poles for $\omega$ and $\omega'$ are different, we see that 
for large $t$, a factor $e^{-2Di(v-u)t}$ oscillating rapidly in time
remains in the integral over $v$; this leads to this integral vanishing in the limit $t \to \infty$.
A similar phenomenon ensures that all terms containing the function $h$ vanish in the same way. Thus, we set $h=0$ in the following, without modifying the results.

\subsubsection{Computing $\Big\langle \delta
g_u \fr{\partial \delta f}{\partial u}\Big\rangle$}
As for $\Big\langle \delta
g_\phi \fr{\partial \delta f}{\partial
\phi}\Big\rangle$, we see that
on the right hand side of equation (\ref{delgudelf}), only the terms
$(k=+1,l=-1)$ and $(k=-1,l=+1)$ are non-zero. We give below the
computation for the case $(k=+1,l=-1)$ case, the other being similar.
Again, we can restrict the computations to the real part of each term, since 
$\Big\langle \delta g_u \fr{\partial \delta f}{\partial u}\Big\rangle$ is real.

From equations (\ref{deltafm1}) and (\ref{deltagu-FT}), we get
\bea
&&\Big\langle \widetilde{\delta
g}_{u,+1}(u,\omega)\fr{\partial \widetilde{\delta
f}_{-1}(u,\omega')}{\partial
u}\Big\rangle\nonumber \\
&&=\fr{\sqrt{1-u^2}}{4i}\fr{\partial }{\partial u}
\Big[\fr{\sqrt{1-u^2}f_0'(u)}{[2Du-m_z[f_0]+\omega']}\langle \widetilde{\delta
m}_-(\omega)\widetilde{\delta
m}_+(\omega')\rangle\Big]\nonumber \\
&&-\fr{1}{2}\sqrt{1-u^2}\fr{\partial }{\partial
u}\Big[\fr{1}{2Du-m_z[f_0]+\omega'}\langle \widetilde{\delta
m}_-(\omega)\delta f_{-1}(u,0)\rangle\Big] \nonumber \\
&&=\frac{\pi}{2i}\fr{\sqrt{1-u^2}}{\epsilon_{+1}(\omega)\epsilon_{-1}(\omega')} \fr{\partial }{\partial u}
\Big[\fr{\sqrt{1-u^2}f_0'(u)}{[2Du-m_z[f_0]+\omega']} \nonumber \\
&&\times \int_{-1}^1 {\rm d}v
\frac{(1-v^2)f_0(v)}{(2Dv-m_z[f_0]-\omega)(2Dv-m_z[f_0]+\omega')}\Big]
\nonumber \\
&&+\fr{i}{2}\fr{\sqrt{1-u^2}}{\epsilon_{+1}(\omega)}\fr{\partial }{\partial
u}\Big[\fr{\sqrt{1-u^2}f_0(u)}{(2Du-m_z[f_0]+\omega')(2Du-m_z[f_0]-\omega)}\Big]
\nonumber \\
&&\equiv a_1+a_2,
\eea
where, in obtaining the second equality, we have used equations (\ref{corr-1}) and (\ref{corr-2}).

Now, to compute the contribution to $\Big\langle \delta
g_u \fr{\partial \delta f}{\partial u}\Big\rangle$ for
($k=+1,l=-1$), we have to integrate $a_1$ and $a_2$ over $\omega$
and $\omega'$. Let us define
\begin{eqnarray}
A_1 \equiv \int_\Gamma \int_{\Gamma'}\frac{{\rm d}\omega
{\rm d}\omega'}{4\pi^2}e^{-i(\omega+\omega')t}a_1, \l{A1}
A_2 \equiv \int_\Gamma \int_{\Gamma'}\frac{{\rm d}\omega
{\rm d}\omega'}{4\pi^2}e^{-i(\omega+\omega')t}a_2. \l{A2}
\end{eqnarray}
We want to compute $A_1$ and $A_2$ in the limit $t\to \infty$, so that we
may discard all terms decaying for large $t$.

Let us first compute $A_1$. Integration over $\omega$ and $\omega'$ may
be carried out by deforming the contours $\Gamma$ and $\Gamma'$, as done
in the preceding subsection. One gets
\bea
\int_\Gamma \frac{{\rm d}\omega}{2\pi}\fr{e^{-i\omega
t}}{\epsilon_{+1}(\omega)(2Dv-m_z[f_0]-\omega)}=i\frac{e^{-i(2Dv-m_z[f_0])t}}{\epsilon_{+1}(2Dv-m_z[f_0])},
\eea
and
\bea
&&\int_{\Gamma'} \frac{{\rm d}\omega'}{2\pi}\fr{e^{-i\omega'
t}}{(2Du-m_z[f_0]+\omega')(2Dv-m_z[f_0]+\omega')\epsilon_{-1}(\omega')} \nonumber \\
&&=-i\Big[\frac{e^{i(2Du-m_z[f_0])t}}{2D(v-u)\epsilon_{-1}(-2Du+m_z[f_0])}-\frac{e^{i(2Dv-m_z[f_0])t}}{2D(v-u)\epsilon_{-1}(-2Dv+m_z[f_0])}\Big],
\eea
so that in the limit $t \to \infty$, we have
\bea
&&A_1=-i\frac{\pi}{2}\sqrt{1-u^2} \fr{\partial }{\partial u}
\Big[\sqrt{1-u^2}f_0'(u) \int_{-1}^1 {\rm d}v
\frac{(1-v^2)f_0(v)}{2D(v-u)\epsilon_{+1}(2Dv-m_z[f_0])}\nonumber \\
&&\times \Big[\frac{e^{-i2D(v-u)t}}{\epsilon_{-1}(-2Du+m_z[f_0])}-\frac{1}{\epsilon_{-1}(-2Dv+m_z[f_0])}\Big]\Big]
\nonumber \\
&&=-\fr{\pi^2}{4D}\sqrt{1-u^2}\fr{\partial }{\partial u}
\Big[\fr{\sqrt{1-u^2}(1-u^2)f_0(u)f_0'(u)}{\epsilon_{+1}(2Du-m_z[f_0])\epsilon_{-1}(-2Du+m_z[f_0])}\Big],
\l{A1-final}
\eea
where, in obtaining the second equality, we have used the lemma (\ref{lemma}).

Now, $A_2$ may be computed along the same lines as done for $B_2$ in
the preceding subsection. One gets
\bea
&&{\rm Re}(A_2)=\frac{i}{2}\sqrt{1-u^2}\fr{\partial}{\partial
u}\Big[\frac{\sqrt{1-u^2}f_0(u)}{\epsilon_{+1}(2Du-m_z[f_0]}\Big]\nonumber
\\
&&=\fr{\pi^2}{4D}\sqrt{1-u^2}\fr{\partial}{\partial
u}\Big[\frac{\sqrt{1-u^2}(1-u^2)f_0(u)f_0'(u)}{\epsilon_{+1}(2Du-m_z[f_0])\epsilon_{-1}(-2Du+m_z[f_0])}\Big],
\l{A2-final}
\eea
where, in obtaining the second equality, we have used equation
(\ref{eps-minus1}).
From equations (\ref{A1-final}) and (\ref{A2-final}), we see that
$A_1+{\rm Re}(A_2)=0$. Similarly, one can show that the real part of the contribution to $\Big\langle \delta
g_u \fr{\partial \delta f}{\partial u}\Big\rangle$ from
($k=-1,l=+1$) also vanishes, while the imaginary
part of the contribution to $\Big\langle \delta
g_u \fr{\partial \delta f}{\partial u}\Big\rangle$ from
($k=-1,l=+1$) must cancel that from ($k=+1,l=-1$). So, we conclude that
for our model,
\be
\Big\langle \delta
g_u \fr{\partial \delta f}{\partial u}\Big\rangle=0.
\l{LB2}
\ee
Combining equations (\ref{LB1}) and (\ref{LB2}), we see that the
Lenard-Balescu operator identically vanishes for our model, which is the announced result.

\section{Example of a Vlasov-stationary state: Relaxation to equilibrium}
\l{examples}
In this section, let us consider as an example of an axisymmetric Vlasov stationary
state $f_0$ a state prepared by sampling independently
for each of the $N$ spins the angle $\phi$ uniformly over $[0,2\pi]$
and the angle $\th$ uniformly over an interval of length $(a+b)$ asymmetric about
$\th=\pi/2$, that is, $\th \in [\pi/2-a:\pi/2+b]$. The corresponding single-spin distribution is
\be
f_0(u,\ph)=\fr{1}{2\pi}p(u),
\l{waterbag}
\ee
with $p(u)$, the distribution for $u$, given by
\be
p(u)=\left\{
\begin{array}{ll}
               \fr{1}{\sin a + \sin b} & \mbox{if $u \in \left[-\sin b,\sin
               a\right]$}, \\
               & \\
               0 & \mbox{otherwise}.
               \end{array}
        \right. \\
\l{pth}
\ee
It is easily verified that this state has the energy
\be
e=\Big(\fr{D}{3}-\fr{1}{8}\Big)(\sin^2 a+\sin^2
b)-\Big(\fr{D}{3}-\fr{1}{4}\Big)\sin a \sin b.
\l{eps}
\ee
Since $m_x[f_0]=m_y[f_0]=0$, we have $g_u^0=0$, and also,
$\partial f_0/\partial \phi=0$; it then follows that the state
(\ref{waterbag}) is stationary under the Vlasov dynamics (\ref{Vlasov}).
Note that we have
\be
m_z[f_0]=\fr{\sin a - \sin b}{2}.
\ee

As mentioned after equation (\ref{dispersion-relation}), the condition ${\rm Im}(\omega_{\rm p}) = 0$ will correspond to the marginal
stability of the state (\ref{waterbag}), so that the zeros
$\omega_{\rm p}$ of the dielectric function lie on the real-$\omega$ axis. Let us denote these zeros as
$\omega^*_{\rm pr}$. From equation (\ref{eps-explicit}), we find
that $\omega^*_{\rm pr}$ satisfies 
\bea
&&1+\pi {\rm
P}\int_{-1}^1 {\rm d}u \fr{(1-u^2)f_0'(u)}{2Du-m_z[f_0]\mp\omega^*_{\rm
pr}}\nonumber \\
&&\pm i\pi^2\left.(1-u^2) f_0'(u)\right|_ {(m_z[f_0]\pm\omega^*_{\rm
pr})/(2D)}=0.
\l{dielectric-marginal}
\eea
Equating the real and the imaginary parts to zero,
we get
\bea
&&1+\pi {\rm
P}\int_{-1}^1 {\rm d}u \fr{(1-u^2)f_0'(u)}{2Du-m_z[f_0]\mp\omega^*_{\rm pr}}=0, \l{marginal-stability-real-part}\\
&&\left.(1-u^2) f_0'(u)\right|_ {(m_z[f_0]\pm\omega^*_{\rm pr})/(2D)}=0.\l{marginal-stability-im-part}
\eea

Now, equation (\ref{waterbag}) gives
\be
f_0'(u)=\fr{1}{2\pi (\sin a+\sin b)}[\delta(u+\sin a)-\delta(u-\sin b)],
\ee
so that we obtain from equation (\ref{marginal-stability-im-part}) that
\bea
\delta\Big(\fr{m_z[f_0]\pm\omega^*_{\rm pr}}{2D}+\sin
a\Big)=\delta\Big(\fr{m_z[f_0]\pm\omega^*_{\rm pr}}{2D}-\sin a\Big),
\eea
implying that 
\be
\omega^*_{\rm pr}=-m_z[f_0].
\l{omegapr}
\ee
Using equation (\ref{omegapr}) in equation
(\ref{marginal-stability-real-part}), we get
\be
4D=\fr{\cos^2 a ~\sin b+\cos^2 b ~\sin a}{\sin a ~\sin b~(\sin a +\sin b)},
\l{Dequation}
\ee
which when combined with equation (\ref{eps}) gives the energy
\bea
&&e=e^*=\fr{A}{B};\nonumber \\
&&A=2 (\sin a-\sin b) ~\cos ^2a+2 \cot a ~\sin ^2b
   ~\cos a+2 \cos b ~\cot b ~\sin ^2a\nonumber \\
   &&+2 \cos ^2b ~(\sin
   b-\sin a)-3 (\sin a-\sin b)^2 (\sin a+\sin b), \nonumber
   \\
&&B=24 (\sin a+\sin b),
\l{estar}
\eea
for which the state (\ref{waterbag}) is a marginally stable stationary
solution of the Vlasov equation (\ref{Vlasov}). For energies $e >e^*$,
such a state is linearly stable under the Vlasov dynamics. On the basis
of our analysis in this paper showing the Lenard-Balescu operator being
identically zero, we expect that for finite
$N$, the state relaxes to Boltzmann-Gibbs equilibrium on a timescale
$\sim N^\delta$, with $\delta > 1$. This was
indeed observed in Ref. \cite{Gupta:2011} for the class of initial states
(\ref{waterbag}) that is non-magnetized, that is, $a=b$; in this case,
combining equations (\ref{Dequation}) and (\ref{estar}), we get
\be
e^*|_{a=b}=\fr{D}{3+12D}.
\ee
 For energies $e<e^*|_{a=b}$, the state being linearly unstable under the Vlasov
 dynamics was seen to relax for finite $N$ to the Boltzmann-Gibbs equilibrium state over a
timescale $\sim \ln N$ \cite{Gupta:2011}. 

\begin{figure}[here!]
\centering
\includegraphics[width=100mm]{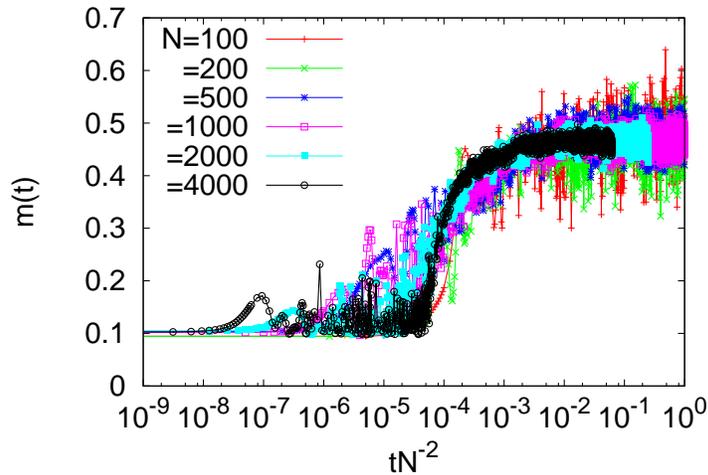}
\caption{
Magnetization $m(t)$ as a function of $tN^{-2}$ in the energy
range in which the state (\ref{waterbag}) with $a=0.1,b=0.3$ is linearly
stable under the Vlasov dynamics. Here, $e=0.22$, $D =8.224$, while the
system sizes are marked in the figure. The 
figure suggests a relaxation timescale $\sim N^{\delta}$ with
$\delta=2$, but the data do not allow a precise determination of the
exponent $\delta$: any value of $\delta$ between about $1.7$ and $2$ is
compatible with the data.}
\l{m-relaxation}
\end{figure}

Let us choose $a=0.1,b=0.3$. Then, equation (\ref{estar}) gives $D
\approx 8.224$, while equation (\ref{estar}) gives $e^*
\approx 0.181$. Thus, for this value of $D$, the state (\ref{waterbag})
with $a=0.1,b=0.3$ is marginally stable under the linearized Vlasov
dynamics at energy $e^* \approx 0.181$. Let us then choose a value of
energy in the range $e^* < e <e_c$, where $e_c$ can be computed from
equation (\ref{ec}) to be $e_c \approx 0.258$. We expect on the basis of
the analysis presented in this paper that in this energy range, when the
state (\ref{waterbag}) is Vlasov-stable, the relaxation to equilibrium
should occur over a timescale that scales superlinearly with $N$. For
$e=0.22$, results of numerical simulations of the dynamics shown in Fig.
\ref{m-relaxation} indeed suggest a relaxation timescale $\sim N^\delta$, with
$\delta > 1$; for the range of system sizes explored in this numerical 
experiment, any value of $\delta$ between about $1.7$ and $2$ is
compatible with the data.

\section{Conclusions}
\l{conclusions}

In this paper, we have shown that the Lenard-Balescu operator
identically vanishes for a system of globally coupled anisotropic
Heisenberg spins, in an axially symmetric Vlasov-stable state. This
result explains the numerical findings of \cite{Gupta:2011}, reporting
a relaxation time for this system that scales superlinearly with
$N$. To our knowledge, it is the first time that this kind of results
has been obtained for a spin dynamics. This raises further questions,
e.g., what are the general conditions to ensure that the Lenard-Balescu
operator vanishes? The classical explanation relies on the 
structure of resonances between the particle trajectories: in the absence 
of resonances between particles with different momentum, the Lenard-Balescu operator should vanish. This heuristic 
argument applies to systems of particles moving in a 1d
position space, thus with a 2d phase space, when the system is homogeneous
\cite{Eldridge:1963,Bouchet:2005R}, implying a relaxation time growing
superlinearly with $N$. This is also the case for
  axisymmetric configurations of point vortices~\cite{Dubin:1988,Chavanis:2012a,Chavanis:2012b}, where the
  phase space is again two-dimensional. In a similar manner, it can be
  argued for the model we have studied that spins with different
  projections on the $z$-axis cannot exchange energy because they cannot be in resonance. In a sense, our precise computations validate this qualitative picture. However, recent numerical
  simulations of a model with a 4d phase space have also shown a
relaxation time that appears superlinear in $N$ over the range of
system sizes studied~\cite{Gupta:2013}: one would expect resonances to 
appear in this case. Thus, understanding the
general conditions under which the Lenard-Balescu operator vanishes
may still remain a partly open question. 

One may also wonder how the relaxation occurs when the Lenard-Balescu operator vanishes. 
Formally, the Klimontovich expansion suggests that the next leading term
is of order $1/N^2$. Although writing down this term is possible in
principle, its evaluation is difficult. However, it is not quite clear that the expansion is valid over such long timescales. 

Finally, let us stress that the standard route to a formal derivation of
the Lenard-Balescu equation, as followed in this article, involves an
averaging over initial conditions. Just as what happens for the Vlasov
equation, one may actually expect that the equation approximately
describes a {\it single} initial condition. Putting this on firm mathematical grounds is an outstanding question, on which some preliminary progress has been made recently \cite{Lancellotti:2009}.

\section{Acknowledgements}
SG acknowledges the support of the Indo-French Centre for the Promotion of Advanced
Research under Project 4604-3 and the hospitality of Laboratoire J. A. Dieudonn\'{e}, Universit\'{e} de
Nice-Sophia Antipolis. 
\vspace{0.25cm}

\end{document}